\documentclass{emulateapj}
\usepackage{graphicx}
\usepackage[usenames]{color}
\usepackage{natbib}

\def \gtw{\>\hbox{\lower.25em\hbox{$\buildrel >\over\sim$}}\>}
\def \ltw{\>\hbox{\lower.25em\hbox{$\buildrel <\over\sim$}}\>}
\def \be{\begin{equation}}
\def \ee{\end{equation}}

\shorttitle{The Crab Pulsar at Centimeter Wavelengths: I}
\shortauthors{Hankins, Jones \& Eilek}    

\begin{document}

\def\ltw{\>\hbox{\lower.25em\hbox{$\buildrel<\over\sim$}}\>}

\title{The Crab Pulsar at Centimeter Wavelengths: I. Ensemble Characteristics}
\author{T.\ H.\ Hankins\altaffilmark{1,2}, G.\ Jones\altaffilmark{3,4}, J.\ A.\ Eilek\altaffilmark{1,2}}
\altaffiltext{1}{Physics Department, New Mexico Tech, Socorro, NM 87801}
\altaffiltext{2}{Adjunct Astronomer, National Radio Astronomy Observatory}
\altaffiltext{3}{Jansky Fellow, National Radio Astronomy Observatory}
\altaffiltext{4}{Physics Department, Columbia University, New York, NY 10027}
\email{thankins@aoc.nrao.edu}

\begin{abstract}
  We have observed the pulsar in the Crab Nebula at high radio frequencies and high time resolution. We present continuously sampled data at 640-ns time resolution, and individual bright pulses recorded at down to 0.25-ns time resolution. Combining our new data with previous data from our group and from the literature shows the dramatic changes in the pulsar's radio emission between low and high radio frequencies. Below about 5 GHz the mean profile is dominated by the bright Main Pulse and Low-Frequency Interpulse.  Everything changes, however, above about 5 GHz; the Main Pulse disappears, the mean profile of the Crab pulsar is dominated by the High-Frequency Interpulse (which is quite different from its low-frequency counterpart) and the two High-Frequency Components. We present detailed observational characteristics of these different components which future models of the pulsar's magnetosphere must explain.
\end{abstract}

\keywords{pulsars: general; pulsars: individual (Crab pulsar)}

\section{Introduction} 
\label{sec:Intro}

Instantaneous pulsar radio emission is strongly variable. Its intensity can fluctuate on timescales as short as microseconds or even nanoseconds. The mean profile, however, is quite stable, and ``components'' of the mean profile are easy to identify. That is, strong radio pulses occur only at certain rotation phases, so that their sum produces a clear component in the time-averaged mean profile. These components therefore represent localized, long-lived ``hot spots'' in the magnetosphere, where plasma conditions are favorable for the creation of coherent radio emission. The fact that the high-energy mean profiles also show distinct components suggests that similar active regions within the magnetosphere create pulses at high photon energies. 

The major question is, where in the magnetosphere are these hot spots? Their locations depend critically on the physics of the magnetosphere: unshielded electric fields, plasma streams, pair creation, plasma and/or MHD instabilities can all contribute to bursts of radio and high-energy emission. Are the hot spots restricted to low altitudes, in the open field line regions above the magnetic polar caps? Do they extend for large distances, out to or even beyond the light cylinder, along high-altitude caustics? Do they exist elsewhere in the magnetosphere, in regions not predicted by any current model?

In this paper we focus on radio emission from the pulsar in the Crab Nebula. Although this star has been widely studied at wavelengths from low radio frequencies through optical, X-rays and $\gamma$-rays, no extant model of that pulsar's magnetosphere can address all of the available data. In this paper we present new observations at high radio frequencies, and use these to characterize the pulsed radio emission from the Crab pulsar in as much detail as possible. We present these results as metrics against which present and future magnetospheric models should be tested.

\subsection{Two Strong Emission Components}

At low radio frequencies (below $\sim 5$ GHz) and at high energies (optical, X-ray, $\gamma$-ray), the mean profile of the Crab pulsar shows two strong features, generally known as the Main Pulse and the Interpulse. The phase separation of the Main Pulse and Interpulse remains approximately constant, $\sim 140^{\circ}$ of rotation phase
(radio, \citet{MH1996};     
optical, \citet{Slowikowska2009}; 
$\gamma$-rays, \citet{Abdo2010}) 
The fact that both radio and high-energy components appear at approximately the same phase strongly suggests the regions emitting both components are located close to each other in the star's magnetosphere. 

\subsubsection{Radio Models}

In models of pulsar radio emission, the Main Pulse and Interpulse are usually interpreted as low-altitude emission from open field lines above the two magnetic poles of an orthogonal rotator, seen at a viewing angle $\sim 90^{\circ}$ between the rotation axis and sight line. Although \citet{Weltevrede2008}  noted there are too few pulsars with interpulses in the observed population to be consistent with a random distribution of viewing angles, the orthogonal-rotator model is supported by polarization observations of some pulsars with interpulses. For instance, \citet{Keith2010} show that the swing of polarization position angle across the Main Pulse and Interpulse in five pulsars with interpulses is consistent with the rotating-vector model \citep{Rad1969}  and a nearly orthogonal viewing angle. However, this model does not easily fit the Crab pulsar, in which the polarization position angle shows no variation with pulse phase across the Main Pulse or the Interpulse \citep{MH1996,Slowikowska2014}.  In addition, the rotating-vector model is further called into question for the Crab pulsar by the shape of  the X-ray torus within the Crab Nebula, which requires a viewing angle $\sim 60^{\circ}$ \citep{NgRom2004}.  Because this angle is too far from orthogonal to allow the low-altitude open field lines from both magnetic poles to pass within our sight line as star rotates, the model of low-altitude polar-cap emission seems not to apply to this pulsar.

\subsubsection{High-Energy Models }
Models of high-energy pulsar light curves paint quite a different picture. Pulsar X-ray and $\gamma$-ray light curves often have two well-defined, sharp peaks separated by $\sim 130^{\circ}$ to $180^{\circ}$ in pulse rotation phase \citep[e.g.,][]{Abdo2013}. Such light curves are too common to be consistent with low-altitude emission from the magnetic poles of orthogonal rotators. In addition, the highest-energy $\gamma$-rays must be emitted from well above the star's surface, because the large pair-creation opacity close to star's surface precludes their escape from low-altitude emission regions \citep[e.g.,][]{RomWat2010}. Models therefore place the emission sites at high altitudes, typically a significant fraction of the distance to the light cylinder, in an outer gap or an extended slot gap. Field line sweepback and photon pile-up on caustics then creates two broad pulses (main pulse and interpulse) when seen at a large range of viewing and inclination angles \citep[e.g.,][]{RomWat2010,Harding2011}.  For the Crab pulsar, the phase coincidence of the Main Pulse and Interpulse in both radio and high energy bands suggests the radio and high-energy emitting regions are regions spatially close in star's magnetosphere. This also seems to disagree with the low-altitude polar cap emission model of radio emission. 

\subsubsection{Alternatives and Complications }
Of course, alternatives to the two-pole, orthogonal rotator model have been suggested. \citet{Manchester1977} argued that that an interpulse is just an extreme version of a two-peaked pulse arising from one magnetic pole. \citet{Hankins1981} pointed out difficulties with the two-pole model of interpulses for PSR B0950+08, and noted that both one-pole and two-pole models are possible for that star. The annular gap model \citep[e.g.,][]{DQW2012}  has both radio and high-energy emission coming from a mid-altitude region above only one pole. Similarly, \citet{RavManH2010} suggested that both radio and high-energy emission in young, fast-spindown pulsars, comes from fan beams emitted at  high altitudes in the magnetosphere. \citet{Pet2009}  proposed that induced Compton scattering of Main Pulse emission creates all other components in the mean profile of the Crab pulsar. 

Furthermore, the Interpulse of the Crab pulsar changes dramatically at high radio frequencies. Above $\sim 5$ GHz, it shifts to $\sim 7^{\circ}$ earlier in rotation phase \citep{MH1996}.  The spectral and temporal nature of single Interpulses is very different from that of single Main Pulses \citep{HE2007}.  We consider the Low-Frequency Interpulse and High-Frequency Interpulse to be two totally separate components.  Even though the two Interpulses are close to each other in phase, and both fall comfortably within the broad high energy Interpulse \citep[e.g.,][]{Abdo2010},  the disparity between the radio characteristics of the two components suggests they arise from different regions within the pulsar's magnetosphere.

\subsection{Even More Radio Components}

This is not yet the entire story of radio emission from the Crab pulsar. Its mean radio profile is complex, with seven separate components identified so far. We list these in order of increasing phase in Table \ref{table:Components}, following the nomenclature of \citet{MH1996}. 

 \begin{deluxetable}{llcc}
\tablecaption{Components of the Mean Profile \label{table:Components}}
\tablehead{
\colhead{Component} & \colhead{Acronym}& \colhead{Frequency} \\
                    &                  & \colhead{Range}     
}
\tablecolumns{3}
\startdata
Precursor                  & PC   & 0.3 - \phn0.6 GHz\\
Main Pulse                 & MP   & 0.3 - \phn4.9\phn GHz\\
High-Frequency Interpulse   & HF IP & 4.2 - 28.4 GHz\\
Low-Frequency Interpulse   & LF IP & 0.3 - \phn3.5 GHz\\
High-Frequency Component 1 & HFC1 & 1.4 - 28.0 GHz\\
High-Frequency Component 2 & HFC2 & 1.4 - 28.0 GHz\\
Low-Frequency Component    & LFC  & 0.6 - \phn4.2 GHz\\
\enddata
\tablecomments{Frequency range over which component is detected in mean profiles \citep[and this work]{MH1996}.  Occasional single pulses may be detected outside this range, but they are too rare to contribute to the mean profile. These acronyms are used only in the Tables and Figures.}
\end{deluxetable}

The Crab pulsar is unusual in having such a large number of components spread throughout the rotation period. While there are many pulsars whose mean profiles contain multiple components \citep[e.g.,  PSR B1237+25 which has five components,][]{Hankins1980} those components are typically bunched closely together in phase and are generally considered part of a single mean-profile component (for instance within the Main Pulse).  By contrast, the seven components found in the Crab pulsar are widely spaced in phase, extending throughout the star's full rotation period.  There are also a few pulsars, such as PSR B0826$-$34 and PSR B1929$+$10, which  emit over most or all of their rotation period.   While some authors  \citep[e.g.,][]{Biggs1985} have suggested the magnetic axis of such stars is oriented close to the line of sight (so-called aligned rotators), this picture is not necessarily consistent with the observed position angle behavior \citep[e.g.,][]{Rankin1997}. 

While the mapping from magnetospheric position to rotation phase is far from linear, the wide phase spread of the Crab's components suggests that many separate radio emission sites exist throughout the star's magnetosphere. We have not seen much discussion in the literature of where in the magnetosphere these components may arise. It has occasionally been suggested that the Precursor is generated at low altitudes near the pulsar surface in the open-field-line region of the polar cap \citep[e.g.,][]{Lyne2013}. This suggestion agrees with caustic models which locate Main Pulse emission, both radio and high-energy, at high altitudes \citep[e.g.,][]{Harding2011}; low-altitude emission arising close to the polar cap  will lead the high-altitude Main Pulse emission by several degrees of phase. Both \citet{HE2007} and \citet{Lyutikov2008} speculated that the High-Frequency Interpulse might arise from high altitudes, close to the light cylinder, but these ideas have not been developed far enough to test against mean profiles. We are not aware of any suggestions for the spatial locations of the other components (the two  High-Frequency Components, and the Low-Frequency Component).  While recent studies of pulsar magnetospheres (e.g., \citet{Li2012, Kalapotharakos2012, Contopoulos2010}) do not specifically model the radio profile, the extended current sheets and dissipation regions revealed by their simulations seem very likely to be sources of coherent radio emission.

\subsection{Our Work: Mean Profiles and Single Pulses}

Our primary focus in this paper is the phenomenology of the mean radio profile of the Crab pulsar. We present continuously sampled data, recorded between 9 and 43 GHz at sub-microsecond time resolution. We use these data to derive and characterize mean profiles at these high radio frequencies, and also identify and characterize the statistics of bright single pulses within the data streams.  In this paper we use the term ``mean profile'' to describe the intensity as a function of rotation phase. This quantity is derived by summing the received intensity synchronously with the star's rotation period. The same data product is also often called the ``light curve'', especially in the high-energy community.

We have also recorded strong individual pulses between 0.33 and 43 GHz, at time resolution down to a fraction of a nanosecond \citep{HE2007, Crossley2010}.  These occasional strong pulses emitted by the Crab pulsar led to its discovery \citep{Staelin1968} long before its periodicity was determined \citep{Comella1969}. These pulses have been called ``giant pulses'', because their intensities are far greater than the typical pulse. However, there is no standard criterion for labeling these single pulses as ``giant'', and in fact the distribution of pulse amplitudes in the Crab pulsar appears to be continuous from weak to strong, \citep[e.g.,][]{Karuppusamy2010}.  In this work we will therefore just refer to ``single pulses'', meaning those bright enough to be clearly discernible above the system noise. Such pulses are relatively rare; in most rotation periods the pulses are below the noise level. When we do detect strong, single pulses, they occur at the phase of the Main Pulse and the Interpulse, and very occasionally at the phases of the High-Frequency Components. 

In this paper we present the ensemble characteristics of the Crab pulsar's radio emission revealed by our data. We describe our observations in Section \ref{section:observations}: we extend up to 43 GHz previous studies at lower radio frequencies, such as that of \citet{Cordes2004}.  In Section \ref{Results:Mean_Profile} we present the composite results: mean profiles as a function of frequency, and the phases and widths of the components of the mean profile. In Section \ref{Results:Single_Pulses} we present ensemble characteristics of individual Mean Pulses and Interpulses: their arrival phase and width, their fluence distributions, and the ``burstiness'' of arrival times. In Section \ref{Summary} we summarize our results and discuss the constraints they place on current and future models of the star's magnetosphere.

In a separate paper to follow, \citep[][Paper 2]{Hankins2015} we extend our previous single-pulse work \citep{Hankins2003, HE2007, Crossley2010} to higher frequencies and more components, by presenting new observations of bright single pulses at the phases of the Main Pulse, the Low-Frequency and High-Frequency Interpulses, and the two High-Frequency Components. 

\begin{figure*}[ht]
\begin{center}
\includegraphics[width=\textwidth]{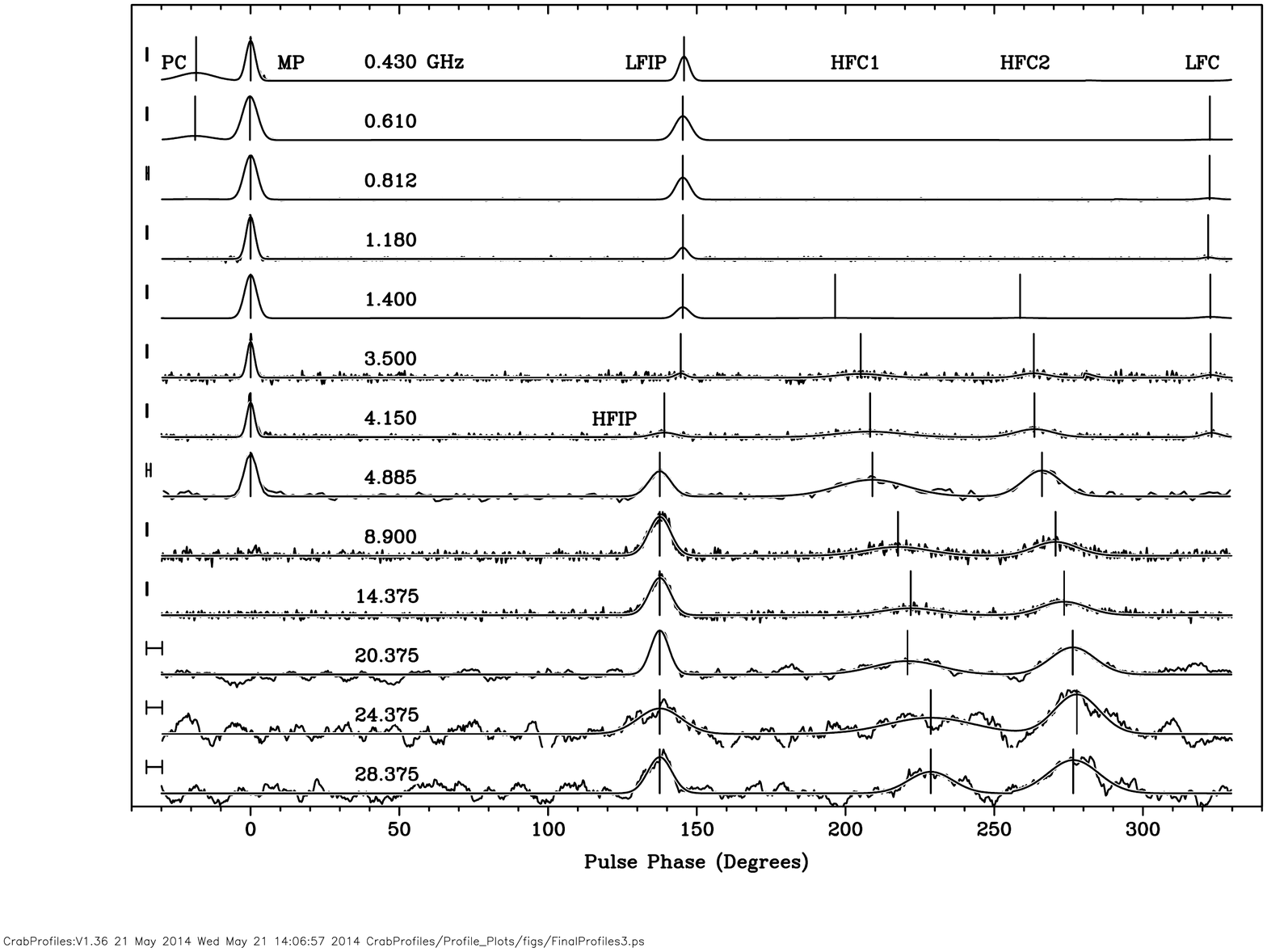}
\caption[]{Mean profiles for a number of frequencies are shown with the formal Gaussian fits to relevant components over-plotted.  The Low-Frequency Component, \emph{LFC}, and the High-Frequency Components, \emph{HFC1} and \emph{HFC2}, were first identified by \citet{MH1996}. The Main Pulse, \emph{MP}, Low-Frequency Interpulse, \emph{LFIP}, and High-Frequency Interpulse, \emph{HFIP}, are also labelled. The 0.43-GHz and 0.61-GHz profiles show the Precursor, \emph{PC}, which only appears below about 1 GHz. The pulse phase of the peaks of the fitted Gaussians are denoted by vertical bars. The  time resolution for each profile is shown by the small horizontal bar at  the left of each profile. 
Sources: 
0.43, 1.18, 3.5, 4.15 GHz: \citet{Cordes2004};
0.61, 1.4 GHz: \citet{Lyne2013};       
0.812 GHz: \citet{Lundgren1995};
4.885 GHz: \citet{MH1999};        
8.9 GHz:  S. Ransom, private communication (2011);        
 $\nu > 10$ GHz: this work.
 \label{fig:FinalProfiles3}}
\end{center}
\end{figure*}

\section{Observations} 
\label{section:observations}

We recorded pulses from the Crab pulsar in observing sessions at the Arecibo Observatory\footnote{The Arecibo Observatory is operated by SRI International under a cooperative agreement with the National Science Foundation (AST-1100968), and in alliance with Ana G. M\'{e}ndez-Universidad Metropolitana, and the Universities Space Research Association.} between 2002 and 2009, the Robert C.~Byrd Green Bank Telescope (GBT)\footnote{The Robert C.~Byrd Green Bank Telescope and the Jansky Very Large Array (JVLA) are instruments of the National Radio Astronomy Observatory, facilities of the National Science Foundation operated under cooperative agreement by Associated Universities, Inc.} in 2009, 2010, and 2011, the Jansky Very Large Array (VLA) from 1993 to 1999.
 
We used two separate data acquisition systems. To obtain continuously sampled data (presented in Sections \ref{Results:Mean_Profile}  and \ref{Results:Single_Pulses}), we used the Green Bank Ultimate Pulsar Processing Instrument (GUPPI) at the GBT, using receiver bands 8-10, 12-15.6, 18-22.4, 22-26.5 and 41-46 GHz and the single linear polarization available at 28-32 GHz.  We operated the GUPPI system in the ``coherent search'' mode, in which the 800-MHz bandwidth was split into 32 frequency bands. Each 25-MHz band was then coherently dedispersed at the nominal Jodrell Bank dispersion measure.\footnote{The Jodrell Bank Crab Pulsar Monthly Ephemeris,  http://www.jb.man.ac.uk/pulsar/crab.html} Total intensity samples were then formed from each time series and every 16 samples were accumulated before being written to disk, resulting in a final time resolution of 640~ns. 
 
To capture individual pulses at higher time resolution (as reported in Section \ref{subsection:Single_Pulse_Widths} and in more detail in Paper 2), we used our Ultra High Time Resolution System (UHTRS) at the Arecibo telescope, the GBT and the VLA.  In this system, a large-memory digital oscilloscope sampled and recorded the received voltages of both polarizations. 
A square-law detector with 20-$\mu$s time constant was used to establish a signal threshold, typically set to 6 times the smoothed off-pulse root-mean-square noise level. When the total intensity exceeded the threshold, the oscilloscope was triggered to sample and digitize the voltages at appropriate Nyquist rates up to 10 Gigasamples per second during a sampling window centered on the pulse component of interest. For the observations at the Arecibo Observatory in 2009 we also used a realtime digital dedisperser based on the CASPER\footnote{Collaboration for Astronomy Signal Processing and Electronics Research, \url{https://casper.berkeley.edu}} iBob device as the oscilloscope trigger generator. The 8-bit oscilloscope-sampled data were transferred to computer disk for subsequent off-line coherent dedispersion \citep{Hankins1971, HR1975},  which allowed time resolution up to the inverse of the receiver bandwidth, about 0.2-0.4 ns. During the data transfer time from oscilloscope to computer disk, 10-30\,s, the data acquisition was disabled, and no pulses could be captured. In Section \ref{subsection:Single_Pulse_Widths} we use data from the UHTRS to characterize widths of single pulses; in Paper 2 we present much more single pulse data taken with the UHTRS.
 
\section{Characteristics of the Mean Profile} 
\label{Results:Mean_Profile}

\subsection{Mean Profiles as a Function of Frequency}
\label{subsection:Mean_profiles}

We have extended the radio mean profiles of \citet{MH1996}  up to 28 GHz in order to test for the frequency dependence of the spacing and width of the mean profile components. In Figure \ref{fig:FinalProfiles3} we show our new profiles aligned with a selection of previously published profiles. As insufficient timing information was available, and because the observations were made at widely separated epochs, we have made two choices to align the profiles. For frequencies $\ltw 5$ GHz, we have set the Main Pulse phase at $0^{\circ}$.  For frequencies $\gtw 5$ GHz we have assumed that the High-Frequency Interpulse phase is fixed at $137.56^{\circ}$ after the center of the Main Pulse. This phase separation is based on several data sets which show a detectable Main Pulse as well as a High-Frequency Interpulse at $\nu > 5$ GHz. Although the \emph{mean profile} of the Main Pulse is very weak above $\sim 5$ GHz,  we have detected occasional single pulses at the phase of the Main Pulse at all frequencies observed, up to 43 GHz (as in Section \ref{subsection:PulseCountHistograms}). 

Figure \ref{fig:FinalProfiles3} shows that the nature of the mean profile changes dramatically between low and high radio frequencies, as follows.

\begin{itemize}

\item The mean profile shape changes abruptly at $\sim 5$ GHz. The Main Pulse, which dominates the mean profile at low radio frequencies, nearly disappears above $\sim 5$ GHz. The high-frequency mean profile is dominated by the Interpulse and the two High-Frequency Components.

\item The Interpulse undergoes a sudden, discontinuous phase shift of $\sim 7^{\circ}$ around 5 GHz. This transition coincides with the frequency above which the spectral and temporal characteristics of the Interpulse change dramatically (Hankins \& Eilek 2007).

\item The two High-Frequency Components can be identified in the mean profile from 1.4 GHz to 28 GHz. Both components drift in phase, appearing at later phases at higher frequencies. 

\item The Precursor can be identified in mean profiles at 0.43 and 0.61 GHz,  but is not seen at any higher frequency. Similarly, the Low-Frequency Component can be seen in mean profiles from 0.61 to 4.2 GHz, but is not seen at higher frequencies.
\end{itemize}

\subsection{Phases of Components in the Mean Profile}
\label{subsection:AvProfCompPhases}

We  simultaneously fitted single Gaussians to each of the profile components in Figure \ref{fig:FinalProfiles3}, ignoring components that are below the profile noise level. The phase of a component is defined as the peak of the fitted Gaussian.  The  phase uncertainties are the quadrature sums of the $1\sigma$ Gaussian centroid positions as obtained from the formal fitting procedure \citep{Press1986}.  The sample point weighting in our least-squares fitting procedure is based on the signal-to-noise ratio, so that  profiles with higher S/N ratios dominate the fits.
\begin{figure}[htb]
{\center
\includegraphics[width=\columnwidth]{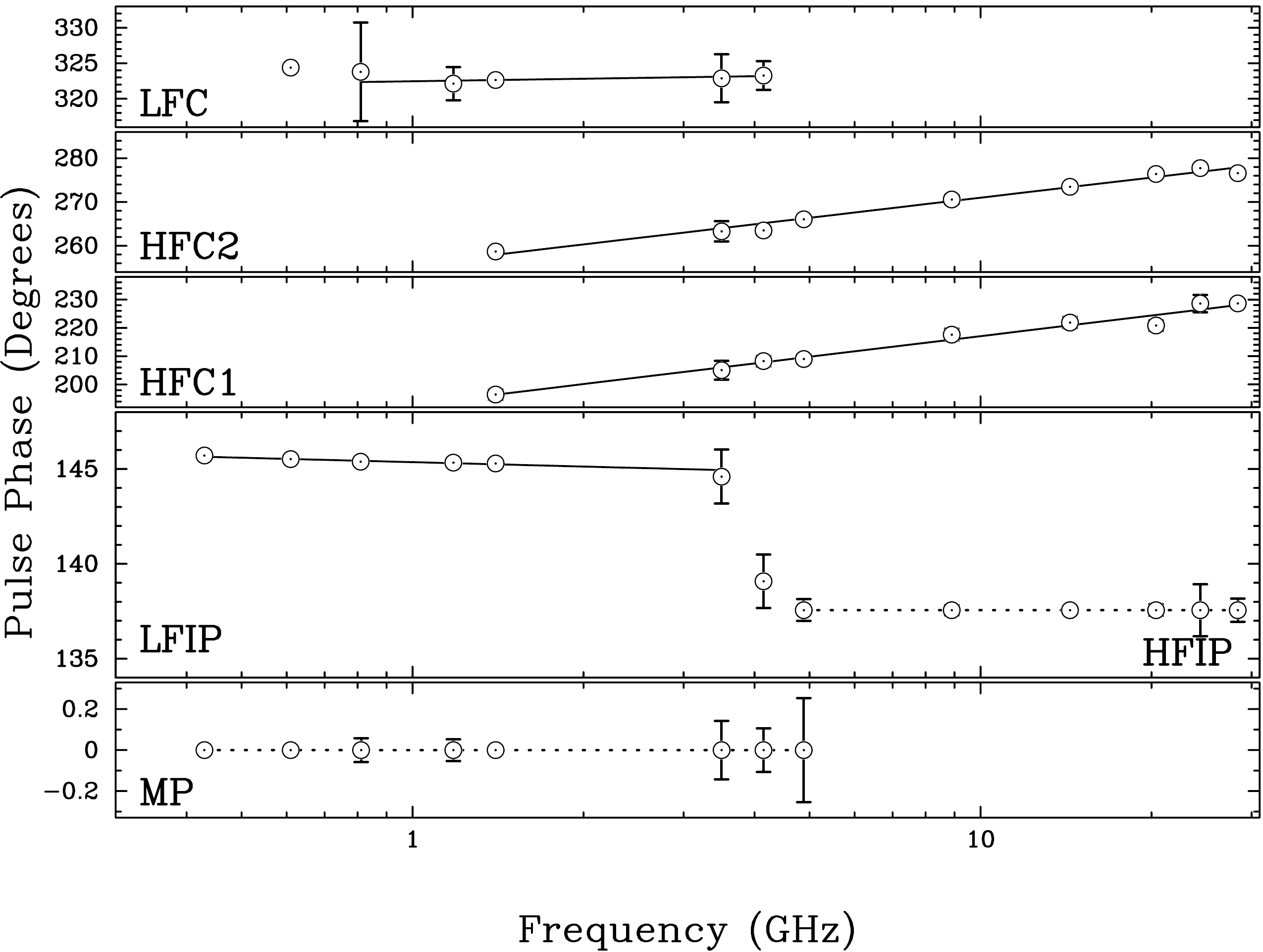}
\caption[]{The phases of the fitted components of the mean profiles. The $1\sigma$ uncertainty error bars for the formal phase fits are not shown when they are smaller than the plot symbol size. The lines are least-squares fits of the form $\phi = a + b\log(\nu)$, where $\phi$ is the pulsar rotational phase in degrees, and $\nu$ is the frequency in GHz. The fit coefficients are given in Table \ref{table:PulsePhases}. The dotted lines indicate that the Main Pulse phase is defined to be zero for $0.4 < \nu < 5$ GHz and the High-Frequency Interpulse phase is defined to be  $137.56^\circ$ for $5 < \nu < 28$ GHz.  We did not include the 0.61-GHz Low-Frequency Component in the formal fit as its position is biased by the stronger, overlapping Precursor.}
\label{fig:HFC_vs_Freq} }
\end{figure}

In Figure \ref{fig:HFC_vs_Freq} we show the phases of each of the fitted components. As in Section \ref{subsection:Mean_profiles}, we assume the Main Pulse has a phase of $\phi_{\rm MP} = 0^\circ$, and the High-Frequency Interpulse has a phase of $\phi_{\rm HFIP} = 137.56^\circ$. For the two High-Frequency Components we computed a fit of phase vs.\ log(frequency), $ \phi(\nu) = a + b \log \nu$, with $\nu$ measured in GHz. We tried fitting polynomials of the form $\phi = c_0 + c_1\nu + c_2\nu^2$ but found that the $\chi^2$ error of the polynomial fit to be much larger. We present the fitting coefficients $a$ and $b$ in Table \ref{table:PulsePhases}, along with the extrapolated values of the phases at a reference frequency of 4 GHz.  Before fitting for  $\phi_{\rm LFIP}(\nu)$ and $\phi_{\rm LFC}(\nu)$ we extrapolated their observed epoch phases to  MJD 53000 using frequency interpolated values of $d\phi/dt$ from \citet{Lyne2013}. The resulting secular phase changes are in all cases smaller than our fitting uncertainties.

\begin{deluxetable}{lccc}
\tablecaption{Mean Profile Component Phases \label{table:PulsePhases}}
\tablehead{
\colhead{Component} & \colhead{$a$} & \colhead{$b$} 
       & \colhead{$\phi(4 ~ {\rm GHz})$} \\
       & \colhead{(degrees)}  & \colhead{(degrees)}  & \colhead{(degrees)}            
}
\tablecolumns{4}
\startdata
MP    &  $\phn\phn0.0\phn\phn\phn\phn\phn\phn\phn$& $0.0\phn\phn\phn\phn\phn\phn$ & $0.0\phn\phn\phn$\\
HF IP &  $137.56\phn\phn\phn\phn\phn\phn$         & $0.0\phn\phn\phn\phn\phn\phn$ & $137.56\phn\phn\phn\phn $\\
LF IP &  $145.36    \pm 0.03     $ & $-0.77   \pm 0.12\phn   $ & $144.9\pm 0.2$\\
HFC1  &  $192.9\phn \pm  0.8\phn $ & $24.2\phn\pm 0.8\phn $    & $207.5\pm 2.5$\\
HFC2  &  $255.7\phn \pm  0.4\phn $ & $15.3\phn\pm 0.4\phn$     & $264.9\pm 1.4$\\
LFC   &  $322.46    \pm 0.09     $ & $\phn1.7\phn\pm 0.5\phn $ & $323.1\pm 0.8$\\[-1pt]
\enddata
\tablecomments{The component phases are fitted by the expression
$\phi(\nu) = a + b \log \nu$, with frequency $\nu$ in GHz. The uncertainties in $a$ and
$b$ are $1 \sigma$; uncertainties in $\phi(4 ~ {\rm GHz})$ are from quadrature sums.  The phases of the Low-Frequency Component and the Low-Frequency Interpulse have been extrapolated to MJD 53000.
}
\end{deluxetable}

These results quantify the trends apparent in Figure \ref{fig:FinalProfiles3}.  The Low-Frequency Interpulse shows very little phase drift with frequency.  The phase separation between the Low-Frequency Interpulse and the Main Pulse, measured from our mean profiles, varies by only $\sim 1.1^{\circ}$ between 430 MHz and 3.5 GHz. Similarly, inspection of our single-pulse data showed no evidence for phase drift with frequency between the Main Pulse and the High-Frequency Interpulse. Except for the $\sim 7^{\circ}$ phase jump when the Low-Frequency Interpulse disappears and the High-Frequency Interpulse appears, each Interpulse remains at a nearly constant phase relative to the Main Pulse.  

\citet{Hankins1986} measured the separation between the Main Pulse and the Interpulse from 0.196 to 2.695 GHz. Their phase data and frequency-dependent fits to the separations are consistent with the current work. They obtained  $\phi_{\rm LFIP}({\rm 1\ GHz}) = 145.9\pm 0.6^{\circ}$ with a frequency exponent of $b = -0.68\pm 0.40$. This compares well with our $\phi_{\rm LFIP}({\rm 1\ GHz}) = 145.36\pm 0.06^{\circ}$ with a frequency exponent of $b=-0.77\pm 0.12$. The secular change of $\phi_{\rm LFIP}$ found by \citet{Lyne2013} over the approximately 20 years from Hankins and Fowler's observations to Lyne's MJD 53000 reference date is far smaller than the quoted separation estimation error of \citet{Hankins1986}.

In the frequency transition region between the Low-Frequency Interpulse and the  High-Frequency Interpulse,  we measured at 4.15 GHz, the Main Pulse to Interpulse separation as $139.1^\circ\pm0.5^\circ$, intermediate between the phases of the Low-Frequency and  High-Frequency Interpulses. Inspection of single-pulse data in this frequency range shows that both types of Interpulses can occur around $\sim 4$ GHz; we thus interpret the intermediate phase of the mean-profile Interpulse at 4.15 GHz as the result of this mixture.

Similarly, neither the Precursor nor the Low-Frequency Component show any strong phase drift over the frequency range in which they appear in our mean profiles. However, the phases of the two High-Frequency Components increase with frequency; at 28 GHz both High-Frequency Components lag their 1.4-GHz counterparts by more than $20^{\circ}$.

The approximate constancy of the Main Pulse-Interpulse phase separation agrees with several other interpulse pulsars \citep[e.g.,][]{Hankins1986}. However, those authors found that the Main Pulse-Interpulse separation in  PSR B1944+17 changes significantly, by $\sim 15^{\circ}$ between 0.43 and 2.4 GHz, reminiscent of the phase drift we find in the High-Frequency Components of the Crab pulsar.  Based on the phase shift in  PSR B1944+17, in addition to polarization anomalies and synchronous nulling in the Main Pulse and Interpulse, \citet{Hankins1986} speculated the radio emission is either from a single magnetic pole or from high altitudes in that star. Might this also be the case for the Crab pulsar?

 \subsection{Widths of Components in Mean Profile}
 \label{subsection:component_widths}

\begin{figure}[htb]
{\center 
\includegraphics[width=\columnwidth]{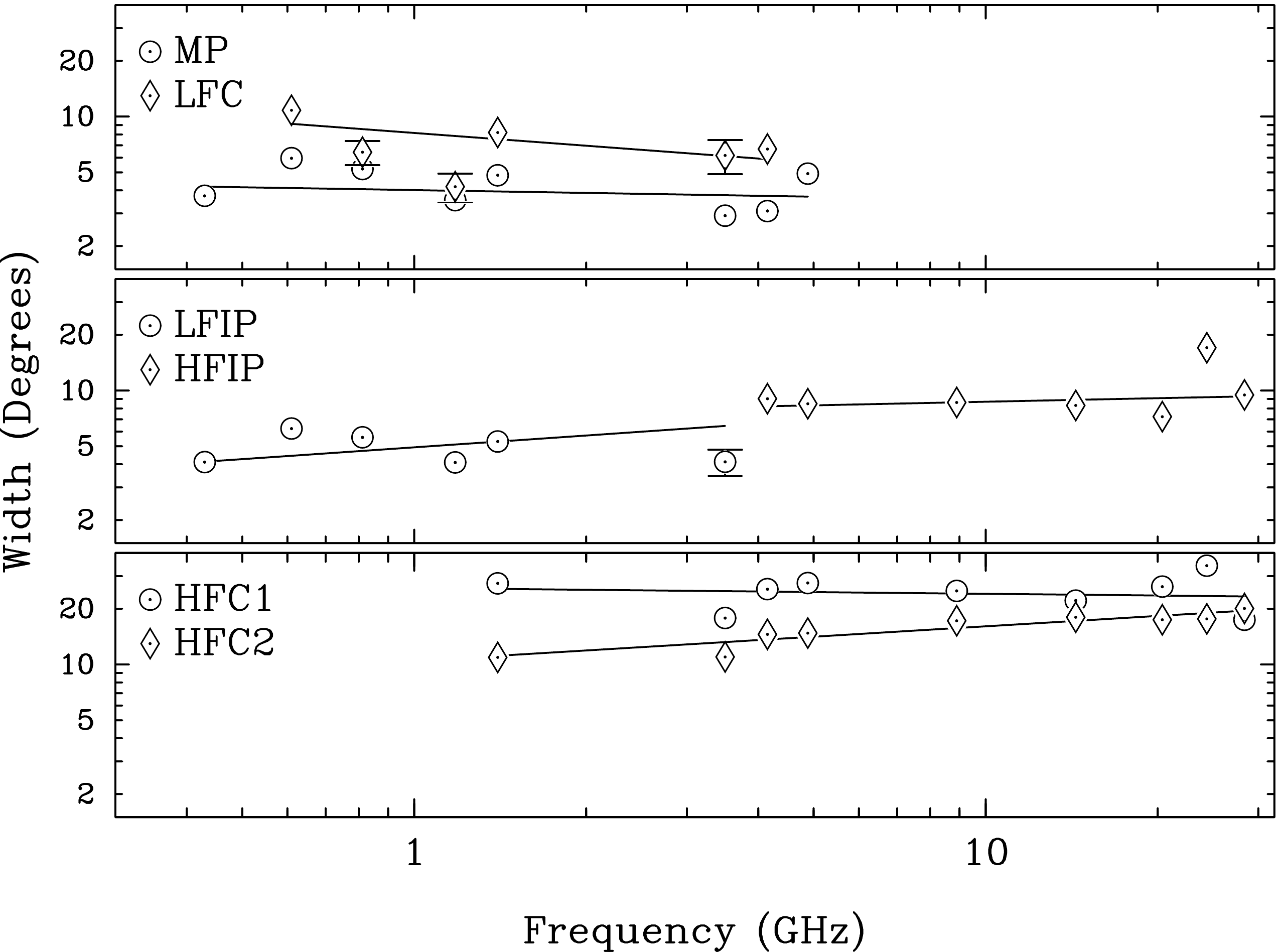}
\caption[]{The full-width at half-maximum of Gaussian fit widths of the mean profiles in Figure \ref{fig:FinalProfiles3} for the Main Pulse, the Low-Frequency and High-Frequency Interpulses,  the two High-Frequency Components, and the Low-Frequency Component.  The fit components are given in Table \ref{table:ComponentWidths}. The error bars are not plotted when the fitted width uncertainty is  less than $10\%$ of the fitted width. For comparison with single-pulse widths (in Figure \ref{fig:Equiv_Widths_vs_frequency_IP_MP}), note that $1^{\circ}$ of phase $\simeq 90\ \mu$s for the Crab pulsar.
\label{fig:Component_Widths}}}
\end{figure}

The Gaussians we fitted to the components in the mean profile for phase determination can also be used to characterize the width of each component. We show in Figure \ref{fig:Component_Widths} the measured full-width at half-maximum of each fitted Gaussian, as a function of frequency. To quantify the relation between width, $w$, and frequency, we fitted a linear function in log space to the data points: $\log w(\nu) =  \alpha + \beta \log\nu$, where $\nu$ is the observing frequency in GHz. We present the fitting coefficients $\alpha$ and $\beta$ in Table \ref{table:ComponentWidths}, along with the widths at 4 GHz, evaluated as $w(4\ {\rm GHz})=10^{\alpha}\nu^{\beta}$. 
  
 The measured widths of the components are, of course, affected by data acquisition systematics, which vary from observation to observation. These include the instrumental time resolution, the dedispersion technique and the value of dispersion measure used, the accuracy of the pulsar ephemeris used to calculate the pulse periods used to form the mean profiles, and, particularly below 1 GHz, the variable scattering broadening in the Crab Nebula and the interstellar medium. 
 An upper estimate of the component width broadened by scattering can be obtained by scaling the values for the scattering decay times, $\tau_{\rm D}$, measured by \citet{Kuzmin2008} at 111 MHz over 2.5 years, as $\nu^{-4}$, and multiplying the result by 2.46 to convert $\tau_{\rm D}$ to $w$, the component full-width at half-maximum.  \citep[Although some authors assume Kolmogorov turbulence for the general ISM, which would imply $\nu^{-4.4}$, the low-frequency pulse broadening of the Crab pulsar is more consistent with Gaussian turbulence; e.\,g.,][]{Crossley2010}. Using the largest value of their observed range, $8 < \tau_{\rm D} < 26$ ms, we get for our lowest observation frequency $w(430\  {\rm MHz}) \approx 2.46 * 26 * (430/111)^{-4} = 0.28$ ms, or about $3.1^\circ$ of pulse phase, which is slightly larger than the size of the plot symbols in Figure \ref{fig:Component_Widths}. Since none of our observations were made during a known scattering event, we conclude that our width measurements are not strongly biased by scattering broadening.

 Despite these uncertainties, we find there is very little change of the component widths over the observed frequency range. The $\beta$ coefficients we find are significantly non-zero only for the Low-Frequency Interpulse and the High-Frequency Component 2, and we note that the formal fit to the Low-Frequency Interpulse  is strongly influenced by two very high S/N points at 0.43 and 1.4 GHz. Thus, only High-Frequency Component 2 shows clear evidence for width changing with frequency.

For many pulsars the mean profile components are broader at low frequencies. For instance, \citet{Rankin1983}  fitted the half-power component widths below $\sim 1$ GHz in several pulsars as $w(\nu) \propto \nu^{\beta}$, with $\beta$ ranging from $-0.5$ to nearly zero. This ``radius-to-frequency mapping'' is often interpreted as lower-frequency radio  emission coming from higher altitudes within the open field line region (but still close to the star's surface).  At higher frequencies, however, component widths tend to stabilize \citep[e.g.,][]{Rankin1983,Thorsett1991}. Our finding of frequency-independent widths for most mean-profile components of the Crab pulsar agree with this general trend. We are not aware of any other pulsar with a mean-profile component that broadens at higher frequencies, as High-Frequency Component 2 does in the Crab.

\begin{deluxetable}{lrrr}
\tablecaption{Mean Profile Component Widths}
\tablehead{
\colhead{Component} & \colhead{$10^{\alpha}$}  
  & \colhead{$\beta$}& \colhead{$w(\rm 4 ~ GHz)$} 
\\
                    & \colhead{(degrees)}&            & \colhead{(degrees)}  
}
\tablecolumns{4}
\startdata
MP    & $\phn4.0\pm 1.1$ & $-0.05 \pm 0.10$ & $\phn3.7 \pm 1.0$ \\
HF IP & $\phn7.5\pm 1.5$ & $ 0.06 \pm 0.15$ & $\phn8.2 \pm 3.1$ \\
LF IP & $\phn4.9\pm 1.0$ & $ 0.21 \pm 0.03$ & $\phn6.6 \pm 1.3$ \\
HFC1  & $   25.9\pm 5.0$ & $-0.03 \pm 0.08$ & $   24.8 \pm 5.0$ \\
HFC2  & $   10.5\pm 1.1$ & $ 0.19 \pm 0.03$ & $   13.5 \pm 1.2$ \\
LFC   & $\phn8.1\pm 1.1$ & $-0.23 \pm 0.19$ & $\phn5.9 \pm 1.9$ \\[-1pt]
\enddata
\tablecomments{The component widths are fitted by the expression $\log w(\nu) = \alpha + \beta \log \nu$, with $\nu$ in GHz. The uncertanties in $\alpha$ and $\beta$ are $1 \sigma$;  uncertainties in $w(4 ~{\rm GHz})$ are from quadrature sum.
\label{table:ComponentWidths}}
\end{deluxetable}

\section{Ensemble characteristics of single pulses}
\label{Results:Single_Pulses}

\begin{figure*}[htb]
\begin{center}
\includegraphics[width=\textwidth]{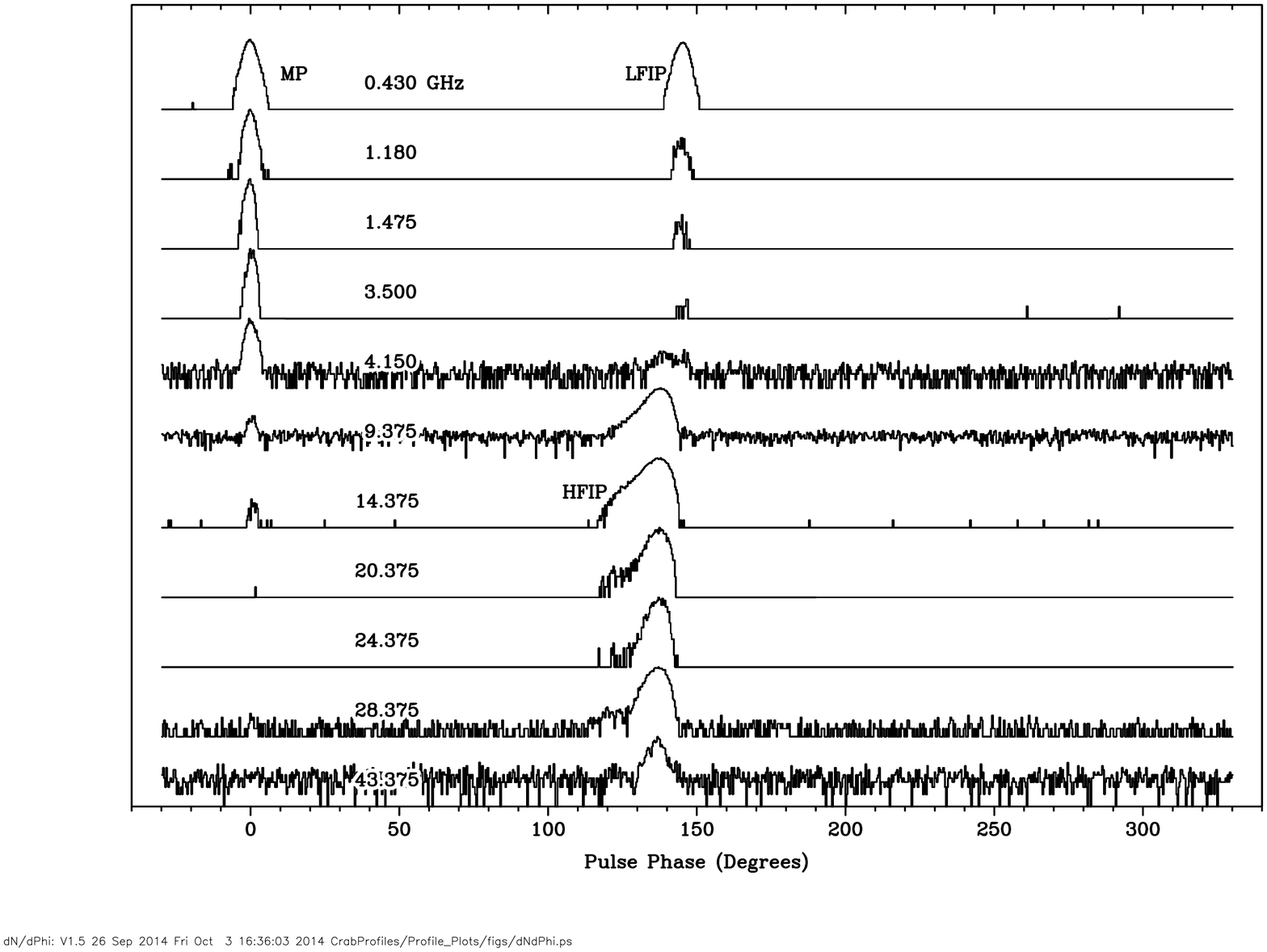}
\caption[]{Phase-resolved logarithmic histograms of the occurrence of single pulses. The histograms have been aligned so that the most probable occurrence of the Main Pulse is at pulse phase zero ($0^{\circ}$). The logarithm of the number of occurrences is normalized to the maximum for plotting. Data sources: $\nu < 5$ GHz, recorded at the Arecibo Telescope \citet{Cordes2004}; $\nu > 5$ GHz, this work, recorded at the GBT.
\label{fig:Pulse_Phase_Histogram}
}
\end{center}
\end{figure*}

\subsection{Single Pulse Arrival Phases}
\label{subsection:PulseCountHistograms}

In addition to the mean intensity profile, we used the GUPPI system to track the occurrence rate of pulses strong enough to be detected as single pulses (above the system noise, typically $6 \sigma$).  In Figure \ref{fig:Pulse_Phase_Histogram} we combine our new data (above 5 GHz) with data from \citet{Cordes2004} to show the frequency evolution of pulse-count histograms of the arrival phase of individual bright pulses. This data product highlights strong, infrequent single pulses which may be hidden in, or distributed differently from, the mean intensity profile (Figure \ref{fig:FinalProfiles3}).

 The histograms in Figure \ref{fig:Pulse_Phase_Histogram} demonstrate that Main Pulses occasionally occur well above 5 GHz, even though they are too infrequent to be detected in the intensity-weighted mean profiles in Figure \ref{fig:FinalProfiles3}. In particular, Main Pulses are evident at 9 and 14 GHz,  and a few were detected at 28 GHz.  Interpulses occur up to 43 GHz, the highest frequency we used. In \citet{HE2007} are shown individual examples of Main Pulses captured with our UHTRS at 9 GHz; in Paper 2 we show examples of single Main Pulses and Interpulses detected above 10 GHz, including one Main Pulse recorded at 43 GHz. We note that both Main Pulses and Interpulses are so rare and weak at 43 GHz that the mean profile does not show either component; thus we did not include 43-GHz data in Figure \ref{fig:FinalProfiles3}. 

The histograms in Figure \ref{fig:Pulse_Phase_Histogram} show no significant detections at the phases of any of the other pulse components. The non-zero histogram values outside the nominal phases of the Main Pulse and Interpulses result from noise spikes or interference that exceeded the signal-to-noise ratio threshold.  

\subsection{Single Pulse Occurrence Rates}
\label{subsection:FreqOccurGPs} 

The pulse-count histograms in Figure \ref{fig:Pulse_Phase_Histogram} clearly show the striking difference between the count rates of the Main Pulse and those of the two Interpulses. At low frequencies the Main Pulse dominates, but above $\sim 5$ GHz---the frequency range where the Interpulse changes nature---the Interpulse dominates the count rates, just as it does the intensity-weighted mean profile in Figure \ref{fig:FinalProfiles3}.

\begin{figure}[htb]
{\center
\includegraphics[width=\columnwidth]{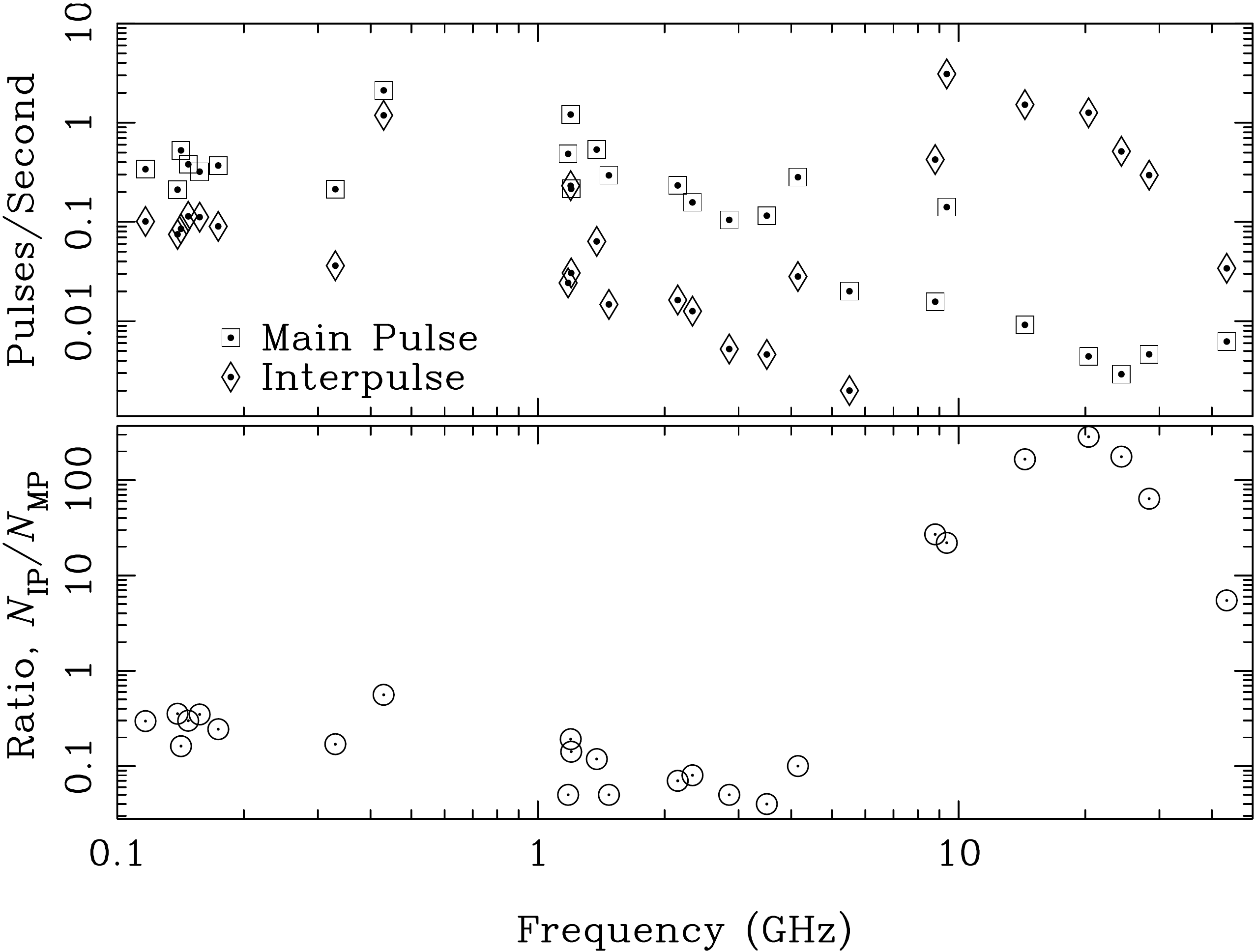}}
\caption{Top: single pulse occurrence rates for Main Pulses, $N_{\rm MP}$, and for Interpulses, $N_{\rm IP}$. Bottom: the rate ratio  $N_{\rm IP}/N_{\rm MP}$, comparing the occurrence rate of Interpulses to the occurrence rate of Main Pulses. Below 5 GHz $N_{\rm IP}$ describes the occurrence rate of Low-Frequency Interpulses; above 5 GHz it describes the occurrence rate of High-Frequency Interpulses.
 Sources: 0.1 to 0.2 GHz, \citet{Karuppusamy2012};
1.197 GHz, \citet{Popov2007};
1.38 GHz, \citet{Karuppusamy2010};
0.43, 1.18, 1.475 to 8.8 GHz, \citet{Cordes2004};
9, 14, 20, 24, 28, 43 GHz, this work.}
\label{fig:IPtoMPratio}
\end{figure}

This can also be shown in terms of count rates, expressed as pulses/second. In Figure \ref{fig:IPtoMPratio} we combine lower-frequency results from the literature with our current high-frequency data,  to show the measured rates of the Main Pulse and two Interpulses, $N_{\rm MP}$ and $N_{\rm IP}$.  Although the measured count rates depend strongly on the observational systematics, such as telescope sensitivity and the Crab Nebula background level, the ratio $N_{\rm IP} / N_{\rm MP}$ is much less sensitive to these effects. 

Figure \ref{fig:IPtoMPratio} shows that bright Main Pulses are somewhat more frequent than bright Low-Frequency Interpulses below $\sim 5$ GHz, but the arrival rates differ only by a factor of a few. However, above $\sim 5$ GHz, the story changes. Bright Main Pulses become increasingly rarer; we typically caught only one every couple of minutes. However,  High-Frequency Interpulses  are much more frequent; we typically detected them 30-1000 times more often than Main Pulses in our continuously sampled observations between 9 and 43 GHz. This statistic, $N_{\rm IP} / N_{\rm MP}$, is somewhat sensitive to the amplitude distributions of the single pulses (discussed below in Section \ref{subsection:AmplitudeDistributions}); but when the rare Main Pulses above 10 GHz are detected, they are certainly no weaker than the High-Frequency Interpulses and sometimes much stronger.

\subsection{Durations of Single Pulses}
\label{subsection:Single_Pulse_Widths}

Pulse-count histograms (Figure \ref{fig:Pulse_Phase_Histogram}) and mean profiles (Figure \ref{fig:FinalProfiles3}) do not tell the full story of the emission physics in the Crab pulsar, because individual pulses emitted at the phase of a given component are generally much narrower than the ensemble width of the component. Our UHTRS observations, obtained at time resolutions from $100\ \mu$s down to $0.25$ ns, allow us to resolve individual bright pulses, and thus characterize their duration, over a wide frequency range. We have recorded single pulses at the VLA  at 0.33, 1.4, 4.8 and 8.4 GHz with 10- to 100-ns time resolution \citep[partial results reported in][]{Crossley2010}; at the Arecibo telescope with frequency bands  4-6, 6-8, and 8-10.5 GHz and 0.4-ns time resolution \citep[partial results reported in][]{HE2007}; and at the GBT with frequency bands 8-10, 12-15.6, 18-22.4, 22-26.5, 28-32 and 41-46 GHz, with 0.25-ns time resolution (as reported in this paper, and in more detail in Paper 2 to follow).

\begin{figure}[htb]
\includegraphics[width=\columnwidth]{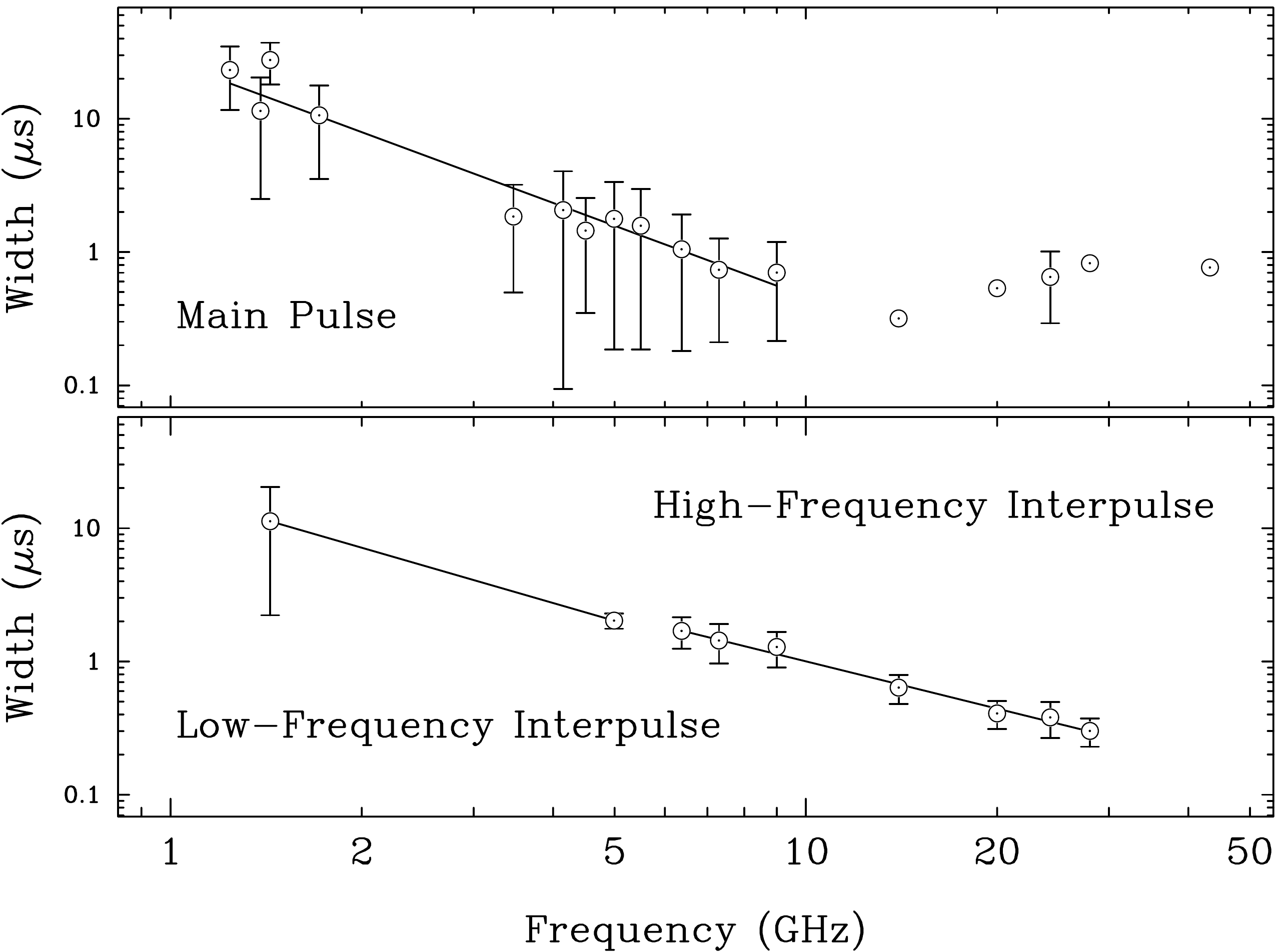}
\caption[]{The averaged equivalent widths of the pulse intensities at various frequencies are shown. The Main Pulse equivalent widths are fitted with a power-law, ${\rm EW}_{\rm MP} \propto \nu^{-1.76\pm0.13}$, between 1.24 and 9 GHz. At each of the frequencies 14, 20, 28 and 43 GHz only one Main Pulse was captured by the UHTRS. For these widths no uncertainty bars are shown. At 24 GHz two Main Pulses were detected. The Low-Frequency Interpulse fit is, ${\rm EW}_{\rm LF\,IP} \propto \nu^{-1.38}$, between 1.4 and 5 GHz. Because this fit has only two degrees of freedom, no error estimate is given. The High-Frequency Interpulse fit is ${\rm EW}_{\rm HF\,IP} \propto \nu^{-1.18\pm0.06}$ between 6.5 and 24.25 GHz. No single High-Frequency Interpulses were detected by the UHTRS above 28 GHz. For comparison with widths of the mean-profile components, note that $1\ \mu$s $\simeq 0.1^{\circ}$ of phase. 
 }
\label{fig:Equiv_Widths_vs_frequency_IP_MP} 
\end{figure}

Because the time signature of individual Main Pulses is quite different from that of individual High-Frequency Interpulses \citep{HE2007}, we need a measure of width that is not overly sensitive to the detailed pulse structure. We choose to calculate the equivalent width (EW) of the pulse intensity.  After experimentation, we found that the EW of the intensity and the EW of the intensity autocorrelation function (ACF) give similar results, differing only by a small multiplicative factor. (The intensity ACF shows a ``zero-lag spike'' due to the noisy structure of the pulse and the contribution of the receiver and sky noise. Therefore, the ACF zero-lag amplitude is strongly dependent upon the signal-to-noise ratio \citep{Rickett1975}.  If used in the calculation of ACF equivalent width, the zero-lag value would strongly bias the width estimate to narrower values. Therefore, for our analysis we truncated the ACF value at zero lag and replaced it by fitting a parabola to adjacent lags before computing the ACF equivalent width.) In Table \ref{table:SinglePulseWidths} we present the number of pulses captured with the UHTRS at each frequency and the mean intensity equivalent width at each frequency. In Figure \ref{fig:Equiv_Widths_vs_frequency_IP_MP} we show the the equivalent width of the pulse intensity against frequency, averaged over all pulses recorded at each frequency. Below 2 GHz our mean equivalent widths for the Main Pulse are consistent with the range of single pulse widths presented in \citet{Crossley2010}  and \citet{Majid2011}.

Comparing these data to the durations of the mean-profile components (Table \ref{table:ComponentWidths} and Figure \ref{fig:Component_Widths}) shows a key result: individual bright pulses are much shorter than the duration of their associated mean-profile component. From Section \ref{subsection:component_widths} we know the width of the Gaussian fit to the  mean profile of the Main Pulse, at a few GHz, is $\sim 370\ \mu$s. For comparison, single bright Main Pulses last at most only $\sim 20\ \mu$s, and usually less, over the same frequency range. Similarly, the width of the High-Frequency Interpulse in the mean profile is $\sim 730\ \mu$s between 5 and 30 GHz, but the width of individual bright Interpulses is no more than $\sim 2\ \mu$s, and usually less, in this frequency range. This result makes it clear that  the mean-profile components do not arise from spatially extended regions with uniform radio emissivity. Rather, the mean profile components are an {\em envelope} inside of which short-lived bursts occur.

Figure \ref{fig:Equiv_Widths_vs_frequency_IP_MP} also shows that the bursts which constitute single pulses become shorter at high frequencies. This is true for Main Pulses, Low-Frequency Interpulses and High-Frequency Interpulses. To quantify this behavior, we carried out least-squares fits of the form ${\rm EW}(\nu) \propto \nu^{\alpha}$. For Main Pulses between 1.24 and 9 GHz, we find $\alpha = -1.76 \pm 0.13$. This fit is approximately consistent with the value of $\alpha \sim -2$ given by \citet{Crossley2010} over a small frequency range (see their Figure 6). For Low-Frequency Interpulses in the range 1.4 to 4.8 GHz we find  $\alpha = -1.38$. For High-Frequency Interpulses in the range 6.5 to 24.25 GHz we find $\alpha = -1.18 \pm 0.06$. This fit is approximately  consistent with the widths reported for single High-Frequency Interpulses by \citet{HE2007} over a much narrower frequency range.

\begin{deluxetable}{crccc}
\tablecaption{Single Pulse Equivalent Widths}
\tablehead{
 \colhead{Frequency} & \colhead{Main Pulse}  & \colhead{Main}   & \colhead{Interpulse}  & \colhead{Inter-}\\
 \colhead{(GHz)}     & \colhead{Width ($\mu$s)}    & \colhead{Pulses} & \colhead{Width ($\mu$s)}    & \colhead{pulses}
 }
\tablecolumns{5}
\startdata
\phn0.333 & $      409.\phn\phn\pm    51.\phn$          &    \phn47 & $ 343.\phn\phn\pm7.\phn\phn$ & \phn29 \\
\phn1.241 & $   \phn23.\phn\phn\pm    12.\phn$          &    \phn32 & & \phn\phn\phn \\
\phn1.385 & $   \phn11.\phn\phn\pm \phn9.\phn$          &    \phn39 & & \phn\phn\phn \\
\phn1.435 & $   \phn28.\phn\phn\pm \phn9.\phn$          &       386 & $  \phn11.\phn\phn\pm9.\phn\phn$ & \phn35 \\
\phn1.714 & $   \phn10.\phn\phn\pm \phn7.\phn$          &    \phn76 & & \phn\phn\phn \\
\phn3.465 & $\phn\phn1.8\phn   \pm \phn1.4$             &    \phn11 & & \phn\phn\phn \\
\phn4.150 & $\phn\phn2.1\phn   \pm \phn2.0$             &    \phn27 & & \phn\phn\phn \\
\phn4.500 & $\phn\phn1.5\phn   \pm \phn1.1$             &    \phn15 & & \phn\phn\phn \\
\phn5.013 & $\phn\phn1.8\phn   \pm \phn1.6$             &    \phn79 & $\phn2.0\phn \pm 0.3\phn$ & \phn\phn3 \\
\phn5.500 & $\phn\phn1.6\phn   \pm \phn1.4$             &    \phn45 &                       & \phn\phn\phn \\
\phn6.500 & $\phn\phn1.05      \pm \phn0.9$             &    \phn43 & $\phn1.7\phn \pm 0.5\phn$ &    \phn33 \\
\phn7.300 & $\phn\phn0.74      \pm \phn0.5$             &    \phn21 & $\phn1.4\phn \pm 0.5\phn$ &    \phn40 \\
\phn9.250 & $\phn\phn0.70      \pm \phn0.5$             &    \phn88 & $\phn1.3\phn \pm 0.4\phn$ &       190 \\
  14.000 & $\phn\phn0.32  \phn\phn\phn\phn\phn\phn$    & \phn\phn1 & $\phn0.64    \pm 0.2\phn$ &    \phn54 \\
  20.000 & $\phn\phn0.53  \phn\phn\phn\phn\phn\phn$    & \phn\phn1 & $\phn0.41    \pm 0.1\phn$ &       130 \\
  24.250 & $\phn\phn0.65  \phn\phn\phn\phn\phn\phn$    & \phn\phn2 & $\phn0.38    \pm 0.1\phn$ &    \phn43 \\
  28.000 & $\phn\phn0.82  \phn\phn\phn\phn\phn\phn$    & \phn\phn1 & $\phn0.30    \pm 0.07$    &    \phn16 \\
  43.250 & $\phn\phn0.77  \phn\phn\phn\phn\phn\phn$    & \phn\phn1 & $\phn\phn\phn\phn\phn\phn\phn$  & \phn\phn\phn \\ %
\vspace{-2ex}
\enddata
\label{table:SinglePulseWidths}
\end{deluxetable}

We also computed the intensity equivalent widths of Main Pulses we captured at 0.33 GHz, finding the average width to be $400\pm50\ \mu{\rm s}$. At this frequency the pulse shape is dominated by interstellar and Nebular scattering broadening \citep[e.g.,][]{Crossley2010}.  We can, however, estimate the intrinsic single-pulse width at this frequency by extrapolating our power-law fit (upper panel of Figure \ref{fig:Equiv_Widths_vs_frequency_IP_MP}) down to 0.33 GHz, which gives $\sim 200\ \mu{\rm s}$. Scaling the typical scattering broadening time of $\tau_{\rm D}(1\ {\rm GHz}) \simeq 2\ \mu{\rm s}$ at 1 GHz \citep[][Figure 6]{Crossley2010} by $\nu^{-4}$ down to 0.33 GHz we obtain $\tau_{\rm D}(0.33\ {\rm GHz}) \simeq 170\ \mu{\rm s}$. If the $200$-$\mu$s intrinsic pulse is convolved with a $170$-$\mu$s scattering function then the expected observed pulse width should be on the order of $\sqrt{200^2 +170^2} \approx 260\ \mu$s, which is not inconsistent with our width measurement.

\subsection{Fluence Distributions}
\label{subsection:AmplitudeDistributions}

\begin{figure}[htb]
\includegraphics[width=\columnwidth]{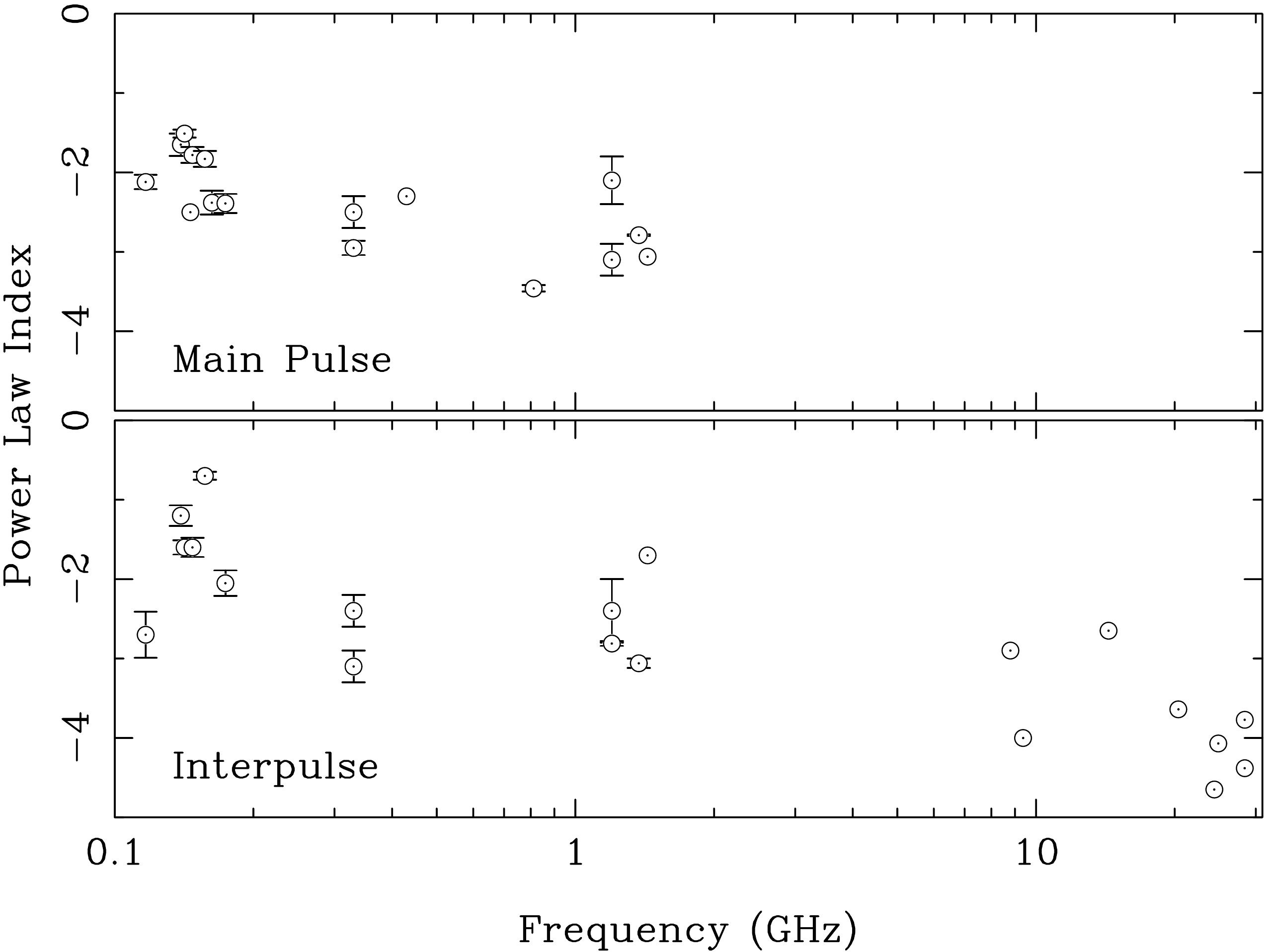}
\caption[]{The power-law indices of single-pulse fluence distributions. The data for $\nu \geq 9$ GHz are for High-Frequency Interpulses measured in this work; other points have been extracted from the literature. Below 5 GHz the power-law index describes the fluence distribution of Low-Frequency Interpulses; above 5 GHz it describes the fluence distribution of High-Frequency Interpulses. Sources for the data points are given in Table \ref{table:FluenceExponents}. Uncertainty error bars are shown when available. }
\label{fig:Indices} 
\end{figure}

\begin{deluxetable}{rllc}
\tablecaption{Power-Law Indices of Fluence Distributions}
\tablehead{
 \colhead{Frequency}     & \colhead{Main Pulse}       & \colhead{Interpulse}   & \colhead{Source} \\
 \colhead{(GHz)}     & \colhead{Index}    & \colhead{Index}& \colhead{}    
}
\startdata
 0.117 &$-2.12   \pm 0.09 $ &$-2.70   \pm 0.29 $ & (a) \\ 
 0.139 &$-1.65   \pm 0.14 $ &$-1.20   \pm 0.13 $ & (a) \\ 
 0.142 &$-1.51   \pm 0.05 $ &$-1.60   \pm 0.09 $ & (a) \\ 
 0.148 &$-1.78   \pm 0.10 $ &$-1.60   \pm 0.12 $ & (a) \\ 
 0.157 &$-1.83   \pm 0.10 $ &$-0.70   \pm 0.05 $ & (a) \\ 
 0.163 &$-2.38   \pm 0.15 $ &                    & (a) \\ 
 0.174 &$-2.39   \pm 0.12 $ &$-2.05   \pm 0.16 $ & (a) \\
 0.146 &$-2.50            $ &                    & (b) \\ 
 0.330 &$-2.50   \pm 0.2  $ &$-2.40   \pm 0.2  $ & (b) \\ 
 0.330 &$-2.95   \pm 0.09 $ &$-3.1\phn\pm 0.2  $ & (c) \\ 
 0.430 &$-2.3\phn         $ &                    & (d) \\ 
 0.812 &$-3.46   \pm 0.04 $ &                    & (e) \\ 
 1.200 &$-2.1\phn\pm 0.3  $ &$-2.4\phn\pm 0.4  $ & (c) \\ 
 1.200 &$-3.1\phn\pm 0.2  $ &$-2.81   \pm 0.03 $ & (c) \\ 
 1.373 &$-2.79   \pm 0.01 $ &$-3.06   \pm 0.06 $ & (f) \\ 
 1.435 &$-3.06   \pm 0.00 $ &$-1.70            $ & (g) \\ 
 8.800 &                    &$-2.9             $ & (d) \\
 9.375 &                    &$-4.00            $ & (h) \\
14.375 &                    &$-2.65            $ & (h) \\    
20.375 &                    &$-3.64            $ & (h) \\     
24.375 &                    &$-4.65            $ & (h) \\     
24.874 &                    &$-4.07            $ & (h) \\   
28.375 &                    &$-5.95            $ & (h) \\    
28.375 &                    &$-4.38            $ & (h) \\    
28.375 &                    &$-3.77            $ & (h) \\    
\vspace{-2ex}
\enddata
\tablecomments{Sources:
(a) \citet{Karuppusamy2012};
(b) \citet{Argyle1972};
(c) \citet{Mickaliger2012};
(d) \citet{Cordes2004};
(e) \citet{Lundgren1995};
(f) \citet{Karuppusamy2012};
(g) \citet{MH1996};
(h) this work.}
\label{table:FluenceExponents}
\end{deluxetable}

We used the continuously sampled GUPPI data to determine the fluence distribution of High-Frequency Interpulses at each frequency we observe between 9 and 43 GHz, averaging over 2-second time bins. (There were not enough Main Pulses in our data to provide reliable statistics). As is conventional, we fitted our fluence distributions with power laws. In Figure \ref{fig:Indices}, also in Table \ref{table:FluenceExponents}, we show the distribution of power-law indices of our data from 9 to 30 GHz along with power-law indices from lower frequency observations from the literature.
\begin{figure*}[htb]
\includegraphics[width=\textwidth]{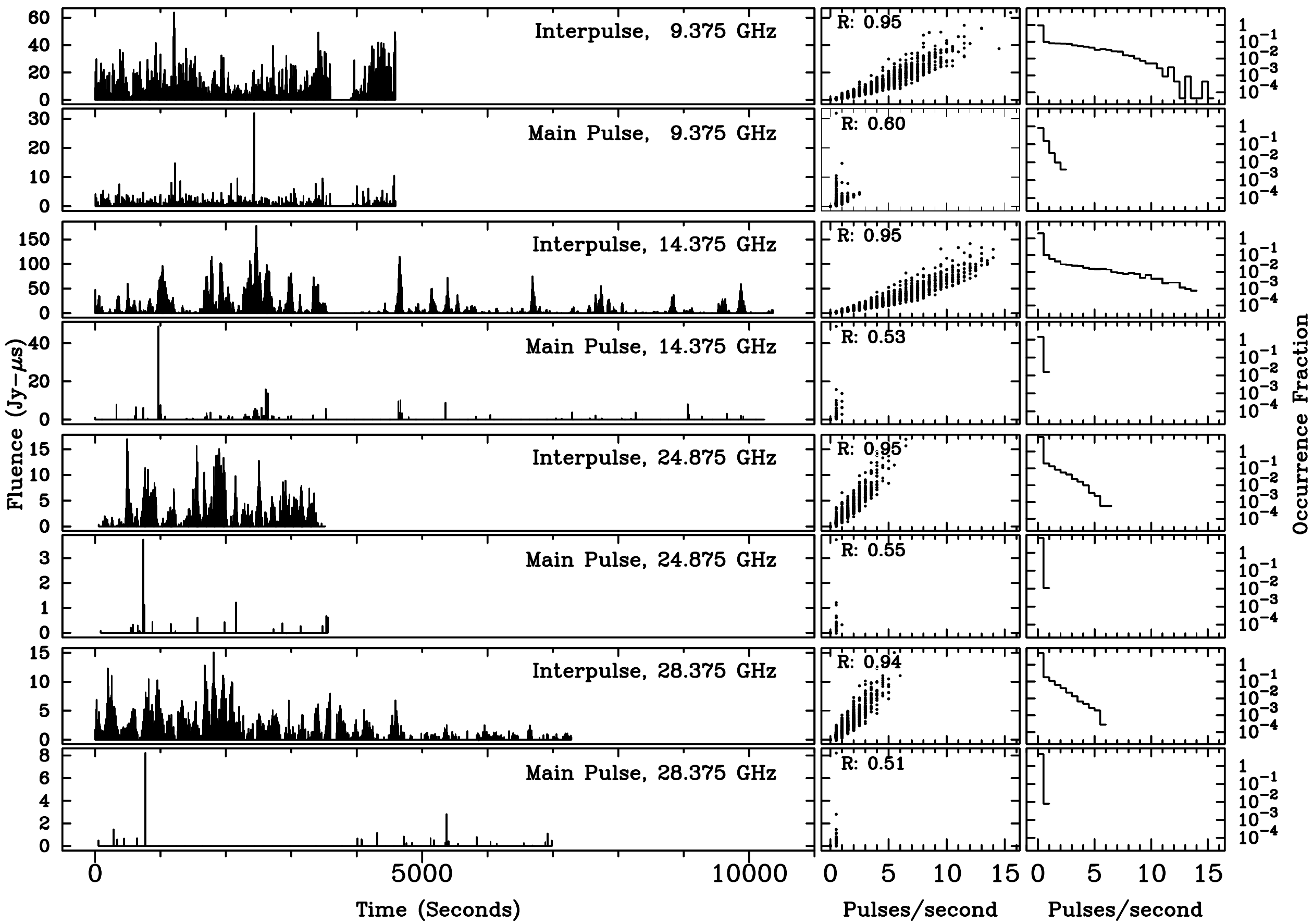}
\caption[]{Occurrence statistics of the Main Pulses and High-Frequency Interpulses for $9 \leq \nu\leq 28$ GHz. The left-hand plots show the 2-second average fluences as a function of time. The central panels show the correlation between fluence and pulse occurrence rates with the correlation coefficient. The right-hand panels show histograms of the fraction of the 2-second intervals during which a particular pulse rate occurred. The rate histogram ordinate exceeds unity for the intervals in which no pulses were detected.}

\label{fig:Occurrence_Rate}
\end{figure*}

There are, of course, many effects that can bias the indices. In addition to different systematic effects which can change with different observations, and the difficulty of establishing true power-law behavior when only a small range of fluence is sampled, the fluence distribution can vary with frequency and between the Main Pulse and the two Interpulses. For example, \citet{Popov2007}, working at 1.197 GHz, found that when the giant pulses are sorted according to their width, the Main Pulse distributions are best fit by two power-law indices, and the Interpulse by one. Also \citet{Karuppusamy2010}, working at 1.38 GHz, found the amplitude distributions are power-law only at the higher intensity tail of the distributions. Therefore, we do not interpret these power-law indices as proof of true power-law behavior of the fluence distribution, but rather as a general indication of the relative abundance of stronger or weaker pulses in that distribution. 

Nonetheless, despite these uncertainties, Figure \ref{fig:Indices} shows that the Main Pulse and both Interpulses share a modest trend. As also noted by \citet{Cordes2004} and \citet{Crossley2010}, there seem to be relatively fewer ``super-giant'' pulses at higher frequencies in each of these components.

\subsection{Bursts of Bright Pulses}
\label{section:Burstiness}

We also used the continuously sampled GUPPI data to study the arrival statistics of bright single pulses, at each observing frequency between 9 and 28 GHz. The different frequencies were not simultaneous, because they were observed on different days, but each Main Pulse/Interpulse time series at a given frequency was extracted from data taken on one observing run. In Figure \ref{fig:Occurrence_Rate} we present three metrics to characterize the arrival statistics on timescales from 2 to 10,000 seconds. 

\subsubsection{The data: Fluctuating Arrival Rates and Fluences } 

In the left panel of Figure \ref{fig:Occurrence_Rate} we show the time series of measured fluences at the phases of the Main Pulse and High-Frequency Interpulse, averaged into 2-second bins. This figure shows that single pulses at these frequencies do not arrive steadily, but rather occur in bursts lasting from a few seconds to several minutes. The separation between bursts is highly variable; sometimes the bursts are separated by only $\sim 100$ seconds, but often it is much longer. There have been observing sessions where we have seen \emph{no} single pulses for several hours. This behavior differs from previous observations at lower frequencies, where pulses are virtually always detectable in any one-second interval (for instance, see Figure \ref{fig:IPtoMPratio}).

The middle panel of Figure \ref{fig:Occurrence_Rate} presents the correlation between the pulse occurrence rates and the pulse fluence. A clear correlation is seen, demonstrating that brighter bursts tend to bunch together. The right panel of Figure \ref{fig:Occurrence_Rate} presents histograms of the pulse arrival rates, characterized as the fraction of the 2-second intervals in which a particular pulse rate was seen. These histograms show that most of the time we recorded none or a only very few pulses per second, but that occasional multi-pulse bursts were possible, especially for the High-Frequency Interpulse which dominates at these high frequencies.

\subsubsection{The Fluctuations: Intrinsic or Interstellar? }

Rapid fluctuations such as seen in the left panel of Figure \ref{fig:Occurrence_Rate} are often ascribed to diffractive interstellar scintillation (DISS) in the strong scattering regime  \citep[e.g.,][]{Cordes1998}.   Because we have the intensity time series over a wide range of wavelengths, we can determine whether DISS is the cause of the burstiness we see between 9 and 28 GHz in the Crab pulsar.

\begin{figure}[htb]
\includegraphics[width=\columnwidth]{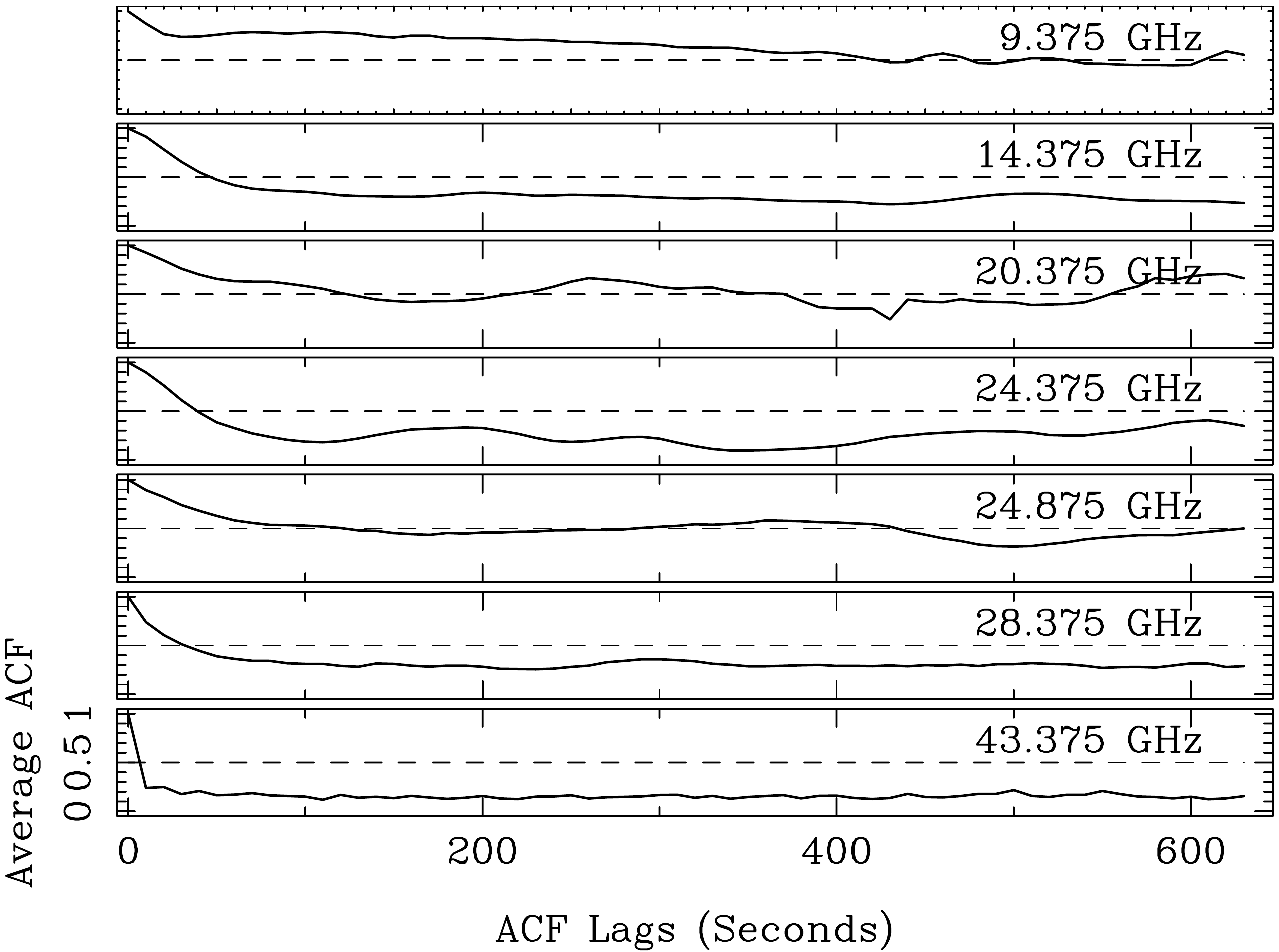}
\caption[]{The normalized autocorrelation of the occurrence rates is shown. The rate values are the number of pulses per two-second interval. Where the ACF crosses the dashed line at 0.5 gives a characteristic time scale of the burstiness.
\label{fig:OccurrenceRateACF}}
\end{figure}

In Figure \ref{fig:OccurrenceRateACF} we show the ACFs of each time series shown in the left panel of Figure \ref{fig:Occurrence_Rate}. The DISS  timescale, $\tau_{{\rm DISS}}$, is typically defined as the half-power point of this ACF \citep[e.g.,][]{Cordes2004}.   Direct inspection of the ACFs in Figure \ref{fig:OccurrenceRateACF} shows that $\tau_{{\rm DISS}}$ fluctuates around $\sim 50$ to 100 s over this range, but does not show any systematic increase with observing frequency $\nu$.  This behavior is inconsistent with standard DISS theory \citep[e.g.,][]{Rickett1990},  which predicts that $\tau_{{\rm DISS}}(\nu) \propto \nu^x$, where $x = 1.0$ for Gaussian turbulence.  Thus, $\tau_{{\rm DISS}}$ should increase by a factor of $\sim 3$ over our 9 to 28 GHz observed frequency range --- which is not the case.  This suggests that the burstiness we observe is intrinsic to the pulsar.

We can also make some simple scaling estimates, following methods in \citet{Cordes1998} also B. Rickett, private communication (2014). The exponential decay constant at 330 MHz, $\tau_{\rm D} \sim 200\ \mu$s, is thought to be due to interstellar scintillation \citep[e.g.,][]{Crossley2010}. If this is the case, we can estimate the interstellar scintillation bandwidth at 330 MHz as $\delta \nu_{{\rm ISS}}(330 ~ {\rm MHz}) = 1 / 2 \pi \tau_{\rm D} \sim 800$ Hz.  This quantity is predicted to increase with frequency, as $\delta \nu_{\rm ISS} \propto \nu^s$, where $s = 2(x+2)$, thus $s=4.0$ for Gaussian turbulence. For instance, we expect $\delta\nu_{\rm DISS} \sim 0.34$ MHz at 1.5 GHz, 670 MHz at 10 GHz, and 55 GHz at an observing frequency of 30 GHz.  At observing frequencies where our receiver bandwidth, $BW$, is comparable to or less than the scintillation bandwidth (e. g., $\delta \nu_{\rm DISS} > 0.2 (BW)$, Cordes \& Rickett 1998),  burstiness such as we observe could be due to interstellar scintillation.  For our 800-MHz GUPPI bandwidth, frequencies $\gtw 7$ GHz could be affected by scintillation.  However, in addition to the fact that our observed $\tau_{\rm DISS}$ does not show the expected linear  increase with frequency, our scaling estimates also suggest the observed burstiness is intrinsic, as follows.

The correlation bandwidth allows us to predict the fluctuation timescale expected from DISS, as $\tau_{\rm DISS}(\nu) \simeq (r_{\rm F} / v) (\delta \nu_{\rm ISS} / \nu)^{1/2}$, where $r_{\rm F} = (c z / 2 \pi \nu)^{1/2}$ is the Fresnel scale, $z$ is the distance to the scattering screen, and $v$ is the speed of the pulsar relative to that screen \citep[e.g.,][]{Cordes1998}. We estimate $v \sim 120$ km/s, from an HST measurement of the pulsar proper motion \citep{Kaplan2007}, predicting $\tau_{\rm DISS} \sim 27$ s at 330 MHz, and $\sim 80$ s at 1 GHz.  For a sanity check, we can compare our expected fluctuation timescale at 1.5 GHz to the results of \citet{Cordes2004}.  These authors formed a time series of giant pulses detected from the Crab pulsar, and measured the $1/e$ width of its ACF as $\sim 25$ s. Identifying this timescale as $\tau_{\rm DISS}$ (which they argued should be detectable at 1.5 GHz with their observing bandwidth), they noted their result is a factor $\sim 3$ shorter than expected from interstellar scintillation. They suggested the discrepancy is due to unusually high speeds of filaments within the Crab nebula; but intrinsic variability from the pulsar on shorter timescales than $\tau_{\rm DISS}$, may be another possibility.

With these scaling arguments, we predict $\tau_{\rm DISS} \sim 800$ s at 10 GHz, and $\sim 2500$ s at 30 GHz. Because these timescales are significantly longer than the burstiness we observed, we believe the burstiness we observe is not due to interstellar scintillation but rather is intrinsic to the radio emission mechanism in the Crab pulsar.

\section{Summary and Discussion}
\label{Summary}

In this paper we have presented new, high-time-resolution observations of the Crab pulsar, at frequencies between 9 and 43 GHz. We combined our new work with previous data taken by our group and from the literature to characterize the mean radio profile and single-pulse statistics of radio emission from the pulsar at these high frequencies. 

\subsection{The Split Personality of the Crab Pulsar} 

Our main result is the dramatic change of the radio emission around frequencies of a few GHz, as follows.

\subsubsection{Low-frequency Radio Profile }
Below $\sim 5$ GHz, the Crab's mean profile is dominated by two bright components, the Main Pulse and the (Low-Frequency) Interpulse. (Two weaker radio components, the Precursor and the Low-Frequency Component can also be detected below $\sim 1$ GHz, close to and leading the Main Pulse in phase.) The Main Pulse and Low-Frequency Interpulse are approximately phase-coincident with the two strong peaks in the pulsar's high-energy profiles. This near-coincidence of emission phases suggests the radio and high-energy emission occur in the same regions of the star's magnetosphere, but that the radio emission sites are more spatially localized (or more tighly beamed) than the high-energy emission regions. 

\subsubsection{High-frequency Radio Profile }
The mean profile is dramatically different above $\sim 5$ GHz; one would hardly believe one is looking at the same star. The Main Pulse disappears almost entirely above $\sim 5$ GHz. The Interpulse --- which is now the High-Frequency Interpulse --- dominates the profile. It appears at a slightly earlier rotation phase than its low-frequency counterpart, but it still sits within the broad profile of the $\gamma$-ray Interpulse. The two High-Frequency Components are also strong in the radio mean profile from $\sim 4$ to 28 GHz. There is no sign of any component in optical, X-ray or $\gamma$-ray mean profiles at the phases of the two High-Frequency Components, except for a possible ($2.3\sigma$) detection of the second High-Frequency Component in the $\gtw 10$ GeV mean profile of \citet{Abdo2010}. 

\subsubsection{Constraints on Future Models}
We do not, unfortunately, have a ready explanation for the geometrical or physical origin of the multiple radio components we see in the Crab pulsar. We therefore close by summarizing the key points about the different mean-profile components in the Crab pulsar, which we hope future models will address.

\subsubsection{How Sudden is the Change? } 
The transition between low and high radio frequencies is sudden, but not ``binary''. Although the Main Pulse disappears from the mean profile above 5 GHz, single bright Main Pulses can still be detected --- and recorded at sub-ns time resolution --- up to 43 GHz.  The change in the Interpulse may be sharper. We have not captured any single Low-Frequency Interpulses above 5 GHz, nor any High-Frequency Interpulses below 4 GHz. We have, however, seen single examples of both Interpulses --- separated by a few degrees of phase --- in UHTRS data taken around 4 GHz. As we suggested in Section  \ref{subsection:AvProfCompPhases} the intermediate phase of the mean-profile Interpulse at 4.15 GHz may reflect a mixture of both types of Interpulses at that frequency.

\subsubsection{Relation of Radio and High-energy Components } 
The Main Pulse and both Interpulses occur at approximately the same phases as their counterparts are seen at optical, X-ray and $\gamma$-rays, which suggests they come from similar regions in the star's magnetosphere. There are, however, important differences. Both the Main Pulse and the Low-Frequency Interpulse are significantly narrower in phase than their high-energy components, and each of these radio components {\em lags} the corresponding high-energy maximum by $\sim200$\,-\,$300\ \mu$s \citep[e.g.,][and references therein]{Abdo2010, Zampiere2014}. Different timing results from different authors and telescopes display some scatter, but all agree that the radio pulses lag the high-energy pulses by a fraction of a ms. The High-Frequency Interpulse also sits within the broad peak of the high-energy Interpulse, and is also narrower than its high-energy counterpart, but it leads the Low-Frequency Interpulse by $7.3^{\circ} \sim 640\ \mu$s, and so it {\em leads} the maximum of the high-energy Interpulse by $\sim 310\ \mu$s. 
 
\subsubsection{Unsteady Radio Emission } 
Radio emission from the Main Pulse and the High-Frequency Interpulse is quite unsteady. Our continuously sampled data show characteristic burst times $\sim30$ to 50 s at all frequencies we recorded between 9 and 28 GHz. We also found that bright bursts can bunch together; and, conversely, the star can go ``quiet'' (no strong radio bursts) for at least several hours. While we cannot definitely prove that the burstiness is not interstellar, the frequency independence of the burst times suggests we are seeing fluctuations intrinsic to the emission process. %

\subsubsection{Mean-profile Components Contain Short-lived Bursts} 
Components of the mean radio profile are built up from the sum of many short-lived radio pulses. Our UHTRS data show that bright single pulses at the phases of the Main Pulse and both Interpulses typically last no more than a few microseconds. This is much shorter than the duration of the mean-profile components, which last several hundreds of microseconds. In addition, we show in Paper 2 that more than one bright single Interpulse can occur during one stellar rotation, within the phase window of the mean-profile component. Thus, a mean-profile component does not represent steady radio emission distributed over an extended region in the magnetosphere; rather, it is the {\em envelope} of where individual pulses can occur.

\subsubsection{Single Pulses at High Time Resolution } 
Much more can be learned by studying single pulses at sub-ns time resolution. In \citet{HE2007} we showed that the High-Frequency Interpulse has different spectral, temporal and dispersion characteristics from the Main Pulse. This suggests that different physical conditions exist in the emission regions, and/or the propagation paths, for each component. In Paper 2 we extend our single pulse analysis to higher radio frequencies and additional components --- and demonstrate that the radio emission physics from the Crab pulsar is even more complex than we have seen up to now.

\begin{acknowledgements}

We thank Jim Cordes, Andrew Lyne and Scott Ransom for the use of their data. We are very grateful to Barney Rickett for illuminating discussions about interstellar scintillation. We also thank the technical, operations, and computer staffs at the GBT the VLA and the Arecibo Observatory for their help with the data acquisition equipment, observing, and for providing some of the computing environment we used for the observations. We are grateful to Jared Crossley, Phil Dooley, Jeff Kern, David Moffett, and James Sheckard for help with the instrumentation, and the observations at Arecibo and the VLA. This work was partially supported by NSF grant AST-0607492. GJ gratefully acknowledges support from a Jansky Fellowship of the National Radio Astronomy Observatory during the course of this work. We thank the referee for helpful comments which have clarified our discussion.
  
\end{acknowledgements}

\end{document}